\documentclass[11pt]{article}

\usepackage{amsfonts,amssymb}
\usepackage{amssymb}
\begin{document}

 \makeatletter \@addtoreset{equation}{section} \makeatother
 \renewcommand{\theequation}{\thesection.\arabic{equation}}
 \baselineskip 15pt

\title{\bf Identical particles and entanglement\footnote{Work supported in
part by Istituto Nazionale di Fisica Nucleare, Sezione di Trieste, Italy}}

\author{GianCarlo Ghirardi\footnote{e-mail: ghirardi@ts.infn.it}\\
{\small Department of Theoretical Physics of the University of
Trieste, and}\\ {\small International Centre
for Theoretical Physics ``Abdus Salam'', and}\\ {\small Istituto
Nazionale di Fisica Nucleare, Sezione di Trieste, Trieste, Italy}\\ and \\
\\ Luca Marinatto\footnote{e-mail: marinatto@ts.infn.it}\\ {\small
Department of Theoretical Physics of the University of Trieste,
and}\\ {\small Istituto Nazionale di Fisica
Nucleare, Sezione di Trieste, Trieste, Italy}}

\date{}

\maketitle

\begin{abstract}
We review two general criteria for deciding whether a pure bipartite quantum
 state describing a system of two identical particles is entangled or not.
The first one considers the possibility of attributing a complete
 set of objective properties to each particle belonging to the composed
 system, while the second is based {\em both} on the consideration of
 the Slater-Schmidt number of the fermionic and bosonic analog of the
 Schmidt decomposition {\em and} on the evaluation of the von Neumann entropy
 of the one-particle reduced statistical operators.
\end{abstract}

\newtheorem{defofentangidentical}{Definition}[section]
\newtheorem{defofentangidentical2}[defofentangidentical]{Definition}
\newtheorem{factorizabilityidentical2}{Theorem}[section]
\newtheorem{factorizabilityidentical3}[factorizabilityidentical2]{Theorem}
\newtheorem{anticomplex}{Theorem}[subsection]
\newtheorem{fermionschmidt}[anticomplex]{Theorem}
\newtheorem{measurefermion}[anticomplex]{Theorem}
\newtheorem{simcomplex}{Theorem}[subsection]
\newtheorem{bosonschmidt}[simcomplex]{Theorem}
\newtheorem{measureboson}[simcomplex]{Theorem}
\newtheorem{bequalb}[simcomplex]{Theorem}
\newtheorem{bdifferentb}[simcomplex]{Theorem}
\newtheorem{entropyin01}[simcomplex]{Theorem}


\section{Introduction}

According to Schr\"odinger quantum entanglement represents
 {\em ``the characteristic trait of Quantum Mechanics, the one that
 enforces its entire departure from classical lines of thoughts''}~\cite{schro}
 due to its peculiar features.
Nowadays entanglement is regarded as the most valuable resource in
 quantum information and quantum computation theory and therefore
 an extensive investigation of its features both from
 the theoretical and from the practical point of view is going on.
In fact the possibility of successfully implementing
 teleportation processes~\cite{tele}, of devising efficient quantum
 algorithms outperforming the classical ones in solving certain computational
 problems~\cite{shor} and of exhibiting secure cryptographical
 protocols~\cite{ekert}, are grounded on the striking physical properties of
 entangled states.
However, despite the fact that almost all the physical realization of
 the above-mentioned processes involve the use of identical particles,
 the very notion of entanglement in systems composed of
 indistinguishable elementary constituents seems to be lacking both of a
 satisfactory theoretical formalization and of a clear physical understanding.
In fact the symmetrization postulate forces the physical systems composed of
 identical fermions and bosons to be described by states possessing
 definite symmetry properties under the permutation of the particle
 indices.
As a consequence, these states generally display (i) a non-factorized
 form, (ii) a Schmidt number greater than $1$, and (iii) a von Neumann
 entropy of the reduced single-particle statistical operator greater
 than $0$~\footnote{The only exception to these statements is represented by
 a system of bosons, each described by the same state vector.}.
Therefore, if one would apply to systems composed of identical constituents
 the criteria which are commonly used for distinguishable particles, one would
 be naturally, but mistakenly, led to the conclusion that non-entangled
 states of identical fermions and bosons cannot exist.

In order to clarify about this common misunderstanding, which originates
 from confusing the unavoidable correlations due to the indistinguishable
 nature of the particles with the genuine correlations
 due to the entanglement, we will briefly and schematically review two
 equivalent criteria we have
 devised for deciding whether a given state is entangled or not.
While the first is based on the possibility of attributing a {\em complete
 set of objective properties} to each component particle of the composed
 quantum system~\cite{gmw,gm}, the second is based {\em both} on the
 consideration of the Slater-Schmidt number of the fermionic and bosonic
 analog of the Schmidt decomposition {\em and} on the evaluation of the
 von Neumann entropy of the one-particle reduced statistical
 operator~\cite{gm2}.


\section{Entanglement for distinguishable particles}

Let us start by recalling the basic features of
 non-entangled (pure) state vectors describing composite systems of two
 distinguishable particles.
Given a bipartite state $\vert \psi(1,2) \rangle \in {\cal H}_{1} \otimes
 {\cal H}_{2}$, the following three equivalent criteria represent
 neccessary and sufficient conditions in order that the state
 can be considered as {\bf non-entangled}:
\begin{enumerate}
\item $\vert \psi(1,2) \rangle$ is factorized, i.e., there
 exist two single-particle states $\vert \phi \rangle_{1} \in {\cal H}_{1}$ and
 $\vert \chi \rangle_{2} \in {\cal H}_{2}$ such that
 $\vert \psi(1,2) \rangle = \vert \phi \rangle_{1} \otimes
 \vert \chi \rangle_{2}$.
In this situation a well-defined state vector is assigned to each component
 subsystem and, since such states
 are simultaneous eigenstates of a complete set of commuting observables,
 it is possible to predict with certainty the measurement outcomes of
 this set of operators.
These outcomes are exactly the objective properties which can be legitimately
 thought as possessed by each particle.
\item The Schmidt number of $\vert \psi(1,2) \rangle$, that is, the number
 of non-zero coefficients appearing in the Schmidt decomposition of the state,
 equals $1$.
\item Given the reduced statistical operator $\rho^{(i)}$
 of one of the two subsystems ($i=1,2$), its von Neumann entropy
 $S(\rho^{(i)}) = - \textrm{Tr}\,[\, \rho^{(i)} \log \rho^{(i)}\,]$ equals
 $0$~\footnote{For our convenience, the $\log$ function is intended
 to be in base $2$ rather than in the natural base $e$.}.
Since the von Neumann entropy measures the uncertainty about the quantum
 state to attribute to a physical system, its value being null mirrors the
 fact that, in this situation, there is no uncertainty at all concerning the
 properties of each subsystem.
\end{enumerate}
On the contrary, a bipartite quantum system is described by an {\bf entangled}
 state $\vert \psi(1,2) \rangle$ if and only if one of the three following
 equivalent conditions holds true: (i) the state is not factorizable; (ii)
 the Schmidt number of the state is strictly greater than $1$; (iii)
 the von Neumann entropy of both reduced statistical operators
 is strictly positive.

In this situation definite state vectors cannot be associated with each
 constituent and therefore we cannot claim that they possess objectively
 a complete set of properties whatsoever.
Accordingly, a strictly positive value of the von Neumann entropy reflects
 this uncertainty concerning the state of the particles.


\section{Two identical particles}

Let us now pass to analyze the case of interest, that is composite
 systems with identical constituents.
In this situation the symmetrization postulate constraints the
 state associated with the system to be totally antisymmetric or
 symmetric under permutation of the two identical fermions or bosons
 respectively.
Consequently the state is no longer factorized, its Schmidt decomposition
 involves generally more than one term and the von Neumann entropy of its
 reduced single-particle statistical operators is strictly
 positive~\footnote{As noticed before
 these statements do not hold true only in the case of two
 bosons which are associated with the same state vector.}.
It is then evident that a bipartite state vector describing
 two indstinguishable particles must (almost always) be considered entangled
 according to the criteria we have outlined in the previous section.
The conclusion we have reached is clearly not correct and
 the origin of the problem resides in not having taken properly into account
 the role played by the unavoidable correlations which are due to
 the indistinguishability of the particles involved, correlations which
 are not connected with those arising from a genuine entanglement.
In order to tackle this problem in the correct way, let us begin
 by sticking to the idea that the physically most interesting and fundamental
 feature of non-entangled states is that {\em both constituents possess a
 complete set of objective properties}.

In Refs.~\cite{gmw,gm} we have taken precisely this attitude, and
 the following definitions formalizing this point have been given:
\begin{defofentangidentical}
 \label{defofentangidentical}
 The identical constituents ${\cal S}_{1}$ and ${\cal S}_{2}$ of a
 composite quantum system ${\cal S}={\cal S}_{1}+{\cal S}_{2}$ are {\bf
 non-entangled} when both constituents possess a complete set of properties.
\end{defofentangidentical}
\begin{defofentangidentical2}
 \label{defofentangidentical2}
 Given a composite quantum system ${\cal S}={\cal S}_{1}+{\cal S}_{2}$ of
 two identical particles described by the normalized state vector
 $\vert \psi(1,2) \rangle$, we will say that one of the constituents
 possesses a complete set
 of properties iff there exists a one-dimensional projection operator $P$,
 defined on the single particle Hilbert space ${\cal H}$, such that:
\begin{equation}
 \label{wittgenstein}
 \langle \psi(1,2)\vert\,{\cal E}_{P}(1,2)\,\vert \psi(1,2)\rangle =1
\end{equation}
\noindent where
\begin{equation}
 \label{operatorediproiezione}
 {\cal E}_{P}(1,2)=P^{(1)}\otimes [\,I^{(2)}-P^{(2)}\,] +
 [\,I^{(1)}-P^{(1)}\,]\otimes  P^{(2)} + P^{(1)}\otimes P^{(2)}.
\end{equation}
\end{defofentangidentical2}
While the first definition has extended to the case of identical
 particles the fundamental feature we have recognised holding
 true for a non-entangled state of two distinguishable particles,
 the second definition is necessary to make precise the meaning
 of the statement {\em ``both constituents possess a complete set of
 properties"} in the considered peculiar situation where it is not possible,
 both conceptually and practically, to distinguish the two particles.
Actually, the condition of Eq.~(\ref{wittgenstein}) gives the probability of
 finding {\it at least} one of the two identical particles (of course,
 one cannot say which one) in the state associated with the
 one-dimensional projection operator $P$.
Since, as already noticed, any state vector is a simultaneous eigenvector
 of a complete set of commuting observables, condition of
 Eq.~(\ref{wittgenstein}) allows to
 attribute to {\em at least} one of the particles the complete set of
 properties (eigenvalues) associated with the considered set of observables.

With the aid of the previous definitions we have been able to
 prove~\cite{gmw,gm} the  following theorems which identify the mathematical
 form displayed by non-entangled state vectors of two identical
 particles:
\begin{factorizabilityidentical2}
 \label{factorizabilityidentical2}
 The identical fermions ${\cal S}_{1}$ and ${\cal S}_{2}$ of a composite
 quantum system ${\cal S}={\cal S}_{1}+{\cal S}_{2}$ described by the
 normalized state $\vert \psi(1,2) \rangle$ are {\bf non-entangled} if and
 only if $\vert \psi(1,2) \rangle$ is obtained by antisymmetrizing a factorized
 state.
\end{factorizabilityidentical2}
\begin{factorizabilityidentical3}
 \label{factorizabilityidentical3}
 The identical bosons of a composite quantum system ${\cal S}={\cal S}_{1}
 + {\cal S}_{2}$ described by the normalized state $\vert \psi(1,2)
 \rangle$ are {\bf non-entangled} if and only if either the state is obtained
 by symmetrizing a factorized product of two orthogonal states or it is
 the product of the same state for the two particles.
\end{factorizabilityidentical3}
These two theorems characterize the cases in which property attribution
 to identical particles is still possible in spite of the non-factorizable
 form of their associated state vectors.
It is necessary to point out that in the situation described by
 Theorems~\ref{factorizabilityidentical2} and~\ref{factorizabilityidentical3},
 not only the property attribution is possible but also
 the peculiar nonlocal correlations between measurement outcomes which are
 typical of the entangled states do not occur.
Accordingly, no Bell's inequality can be violated and no teleportation process
 can be performed~\cite{gmw,gm} by means of a state where each particle still
 possess a definite state vector.


\section{Another criterion for detecting entanglement}

Recently in the scientific literature new criteria for detecting
 entanglement appeared~\cite{cirac,pask,li,vacca}.
Part of them simply consists in a (careless) extension to the case of identical
 particles of the same criteria used when dealing
 with distinguishable particles.
Unfortunately those criteria presents some obscure aspects and sometimes
 they fail to identify certain kinds of non-entangled states:
 in fact while some of them~\cite{cirac,pask} correctly deal with the case of
 identical fermions, an inappropriate treatment of the (subtle) boson case is
 presented in Ref.~\cite{pask} and in Ref.~\cite{li} the use of the entropy
 criterion  is misleadig.
With the aim of overcoming such puzzling situations,
 we have presented an unambiguous criterion to identify whether
 a state is entangled or not, which resort simultaneously
 {\em both} to the consideration of
 the Slater-Schmidt number of the fermionic and bosonic analog of the
 Schmidt decomposition {\em and} to the evaluation of the von Neumann entropy
 of the one-particle reduced statistical operators.

Such a criterion completely agrees with the criterion based on the property
 attribution which we reviewed in the previous section and
 it seems to settle all the puzzling issues which have been pointed out by
 the authors of Refs.~\cite{pask,li,vacca}.

We present such a criterion dealing first with the simpler case of two
 identical fermions and, subsequently, we pass to analyze the
 more subtle case of two identical bosons.


\subsection{The fermion case}

The notion of entanglement for systems composed of two identical
 fermions has been discussed in Ref.~\cite{cirac} where a
 {\em ``fermionic analog of the Schmidt decomposition''} was exhibited.
This decomposition derives from an extension to the set of the
 antisymmetric complex matrices of a well-known theorem holding
 for antisymmetric real matrices and it states that:
\begin{fermionschmidt}
 \label{fermionschmidt}
 Any state vector $\vert \psi(1,2)
 \rangle$ describing two identical fermions of spin $s$ and,
 consequently, belonging to the antisymmetric manifold
 ${\cal A}(\mathbb{C}^{2s+1}\otimes \mathbb{C}^{2s+1})$, can be written as:
\begin{equation}
 \label{fermion0.5}
 \vert \psi(1,2)\rangle = \sum_{i=1}^{(2s+1)/2} a_{i}\cdot
 \frac{1}{\sqrt{2}} \,[\,\vert 2i-1\rangle_{1} \otimes \vert 2i
 \rangle_{2}- \vert 2i \rangle_{1} \otimes \vert 2i-1 \rangle_{2}\,],
\end{equation}

\noindent where the states
 $\left\{\, \vert 2i-1 \rangle, \vert 2i \rangle \right\}$
 with $i=1\dots (2s+1)/2$ constitute an orthonormal basis of
 $\mathbb{C}^{2s+1}$, and the complex coefficients $a_{i}$ (some of which may
 vanish) satisfy the normalization condition $\sum_{i} \vert a_{i}\vert^{2}=
 1$.
\end{fermionschmidt}
The number of non-zero coefficients involved in the decomposition
 of Eq.~(\ref{fermion0.5}) is called the {\em Slater numer} of the state
 $\vert \psi(1,2)\rangle$.
We distinguish two cases:\\

\noindent {\bf Slater Number $=1$}. In this situation the state
 $\vert \psi(1,2)\rangle$ has the form of a single Slater determinant:
\begin{equation}
\label{fermion1}
 \vert \psi(1,2)\rangle = \frac{1}{\sqrt{2}}\,[\vert 1\rangle_{1}\otimes
 \vert 2\rangle_{2} - \vert 2\rangle_{1}\otimes \vert 1\rangle_{2}\,]
\end{equation}

\noindent Since the state has been obtained by antisymmetrizing the product of
 two orthogonal states, $\vert 1\rangle$ and $\vert 2 \rangle$, it must
 be considered as non-entangled according to
 Theorem~\ref{factorizabilityidentical2}.
The reduced single-particle statistical operators of each particle
 (it does not really matter
 which one we consider since, due to symmetry considerations, they
 are equal) and their von Neumann entropy (expressed in base $2$) are:
\begin{equation}
\label{fermion2}
\rho^{(1\:or\:2)} = \frac{1}{2}\,[\, \vert 1 \rangle \langle 1
 \vert + \vert 2 \rangle\langle 2 \vert \,]
\end{equation}
\begin{equation}
\label{fermion3}
 S(\rho^{(1\:or\:2)}) \equiv - \textrm{Tr} \,[\, \rho^{(1\:or\:2)}
 \log \rho^{(1\:or \:2)}\,] = 1
\end{equation}

In this situation, the value $S(\rho^{(1\:or\:2)})=1$
 correctly measures only the unavoidable uncertainty concerning the quantum
 state to attribute to each of the two identical physical subsystems and
 it has nothing to do with any uncertainty arising from any actual
 form of entanglement.
In fact it should be obvious that we cannot pretend that the operator
 $\rho^{(1\:or\:2)}$ of Eq.~(\ref{fermion2}) describes the properties of
 {\em precisely} the first or of the second particle of the system,
 due to their indistinguishability.\\

\noindent {\bf Slater number $>1$}. In this case the state
 $\vert \psi(1,2)\rangle$ cannot be obtained by antisymmetrizing
 the tensor product of two orthogonal states and, consequently, its
 decomposition involves more than one single Slater determinant.
Therefore, as a consequence of the criterion expressed by
 Theorem~\ref{factorizabilityidentical2}, the state $\vert \psi(1,2)
 \rangle$ must be considered as a truly entangled state.
The reduced single-particle statistical operators and their
 associated von Neumamn entropy are:
\begin{equation}
\label{fermion4}
\rho^{(1\:or\:2)} = \sum_{i=1}^{(2s+1)/2} \frac{\vert a_{i}\vert^{2}}{2}
 \,[  \vert 2i-1 \rangle \langle 2i-1 \vert + \vert 2i \rangle \langle
 2i\vert\,]
\end{equation}
\begin{equation}
\label{fermion5}
 S(\rho^{(1\:or\:2)}) = -\sum_{i=1}^{(2s+1)/2} \vert a_{i}\vert^{2}
 \log \frac{\vert a_{i}\vert^{2}}{2} =
 1-\sum_{i=1}^{(2s+1)/2} \vert a_{i}\vert^{2} \log \vert a_{i}\vert^{2} > 1
\end{equation}

In this case the von Neumann entropy is strictly greater than $1$ and
 it correctly measures both the uncertainty deriving from the
 indistinguishability of the particles and the one connected with the genuine
 entanglement of the state. \\

\noindent The previous two cases can be summarized in the following theorem:
\begin{measurefermion}
 \label{measurefermion}
 A state vector $\vert \psi(1,2) \rangle$ describing two identical
 fermions is {\bf non-entangled} if and only if its Slater number is equal
 to $1$ or equivalently if and only if the von Neumann entropy of the
 one-particle reduced statistical operator $S(\rho^{(1\: or \:2)})$ is equal
 to $1$.
\end{measurefermion}


\subsection{The boson case}

Let us pass now to analyze the case of bipartite quantum systems composed of
 two identical bosons.
This case turns out to be slightly more articulated and
 subtle than the fermionic case, as a consequence of the peculiar properties
 of the bosonic statistics.
We begin by considering the bosonic Schmidt
 decomposition of an arbitrary state vector $\vert \psi(1,2) \rangle$
 belonging to the symmetric manifold ${\cal S}(\mathbb{C}^{2s+1}\otimes
 \mathbb{C}^{2s+1})$ and describing two identical bosons:
\begin{bosonschmidt}
 \label{bosonschmidt}
 Any state vector
 describing two identical $s$-spin boson particles $\vert \psi(1,2)
 \rangle$ and, consequently, belonging to the symmetric manifold
 ${\cal S}(\mathbb{C}^{2s+1}\otimes \mathbb{C}^{2s+1})$ can be written
 as~\footnote{It is worth pointing out that the Schmidt decomposition of
 Eq.~(\ref{boson1}) is not always unique, as happens for the biorthonormal
 decomposition of states describing distinguishable particles. However,
 the number of non-zero coefficients is uniquely determined.}:
\begin{equation}
 \label{boson1}
 \vert \psi(1,2)\rangle = \sum_{i=1}^{2s+1} b_{i}\,
 \vert i \rangle_{1} \otimes \vert i \rangle_{2}\:,
\end{equation}
where the states $\left\{\, \vert i \rangle \right\}$, with
 $i=1,\dots, 2s+1$, constitute an orthonormal basis for $\mathbb{C}^{2s+1}$,
 and the real nonnegative coefficients $b_{i}$ satisfy the normalization
 condition $\sum_{i}b_{i}^{2}= 1$.
\end{bosonschmidt}
The number of non-zero coefficients $b_{i}$ appearing in the decomposition
 of Eq.~(\ref{boson1}) is called the Schmidt number of the state
 $\vert \psi(1,2)\rangle$.
Then the following cases can occur:\\

\noindent {\bf Schmidt number $=1$}. In this case the state is factorized,
 i.e.
 $\vert \psi(1,2) \rangle= \vert i^{\star} \rangle \otimes \vert i^{\star}
 \rangle$, and it describes two identical bosons in the same state
 $\vert i^{\star} \rangle$.
It is evident that such a state must be considered as non-entangled since
 one knows precisely the properties objectively possessed by each particle
 and, consequently, there is no uncertainty about which particle has which
 property.
This fact perfectly agrees with the von Neumann entropy of the single-particle
 reduced statistical operators $S(\rho^{(1 \,or\,2)})$ being null. \\

\noindent{\bf Schmidt number $=2$}. According to Eq.~(\ref{boson1}), the most
 general state with Schmidt number equal to $2$ has the following form:
\begin{equation}
 \label{boson4.3}
 \vert \psi(1,2) \rangle= b_{1}\vert 1 \rangle_{1} \otimes \vert 1\rangle_{2}
 + b_{2}\vert 2 \rangle_{1} \otimes \vert 2\rangle_{2},
\end{equation}
where $b_{1}^{2}+b_{2}^{2}=1$.

Now two subcases, depending on the values of the positive coefficients
 $b_{1}$ and $b_{2}$, must be distinguished and separately analyzed.
If they are equal, that is, if $b_{1}\!=\!b_{2}=\!1/\sqrt{2}$, the following
 theorem holds:
\begin{bequalb}
 \label{bequalb}
The condition $b_{1}=b_{2}=1/\sqrt{2}$ is necessary and sufficient in order
 that the state $\vert \psi(1,2) \rangle= b_{1}\vert 1 \rangle_{1} \otimes
 \vert 1\rangle_{2} + b_{2}\vert 2 \rangle_{1} \otimes \vert 2\rangle_{2}$
 can be obtained by symmetrizing the factorized product of two orthogonal
 states.
\end{bequalb}
In this situation, and in full accordance with
 Theorem~\ref{factorizabilityidentical3}, one must consider this state as
 non-entangled since it is possible to attribute definite state vectors
 (and consequently definite objective properties) to both particles.
As usual, we cannot say which particle is associated with which state due to
 their indistinguishability.
Moreover the von Neumann entropy of the reduced statistical operators
 $S(\rho^{(1\:or\:2)})$ is equal to $1$ and it measures only the uncertainty
 descending from the indistinguishability of the particles, as happened
 in the fermion case with the state of Eq.~(\ref{fermion1}),

On the contrary, when the two coefficients are different, that is,
 when $b_{1}\neq b_{2}$, the following Theorem holds:
\begin{bdifferentb}
 \label{bdifferentb}
The condition $b_{1}\neq b_{2}$ is necessary and sufficient in order that
 the state $\vert \psi(1,2) \rangle= b_{1}\vert 1 \rangle_{1} \otimes
 \vert 1\rangle_{2} + b_{2}\vert 2 \rangle_{1} \otimes \vert 2\rangle_{2}$
 can be obtained by symmetrizing the factorized product of two
 non-orthogonal states.
\end{bdifferentb}
According to our original criterion, this state must be considered
 as a truly entangled state since it is impossible to attribute to both
 particles definite state vectors (and, consequently, definite
 objective properties).
In this situation the von Neumann entropy of the reduced statistical operator
 $S(\rho^{(1\:or\:2)})= -b^{2}_{1}\log b_{1}^{2} - b_{2}^{2}\log b_{2}^{2}$
 lies within the open interval $(0,1)$.
It correctly measures simultaneously the uncertainty arising both from the
 indistinguishability of the particles and from the entanglement.
It is strictly less than $1$ because, in a measurement process, there is a
 probability greater than $1/2$ to find both bosons in the
 same physical state ($\vert 1\rangle$ or $\vert 2 \rangle$ depending
 whether $b_{1}>b_{2}$ or vice versa).\\

\noindent {\bf Schmidt number $\geq 3$}. In this situation the state is a
 genuine entangled one since
 it cannot be obtained by symmetrizing a factorized product of two
 orthogonal states~\footnote{In fact, if this would be true, the rank of the
 reduced density operator would be equal to $2$, in contradiction with the
 fact that a Schmidt number greater than or equal to $3$ implies a rank equal
 to or greater than $3$.} and the von Neumann entropy
 of the reduced statistical operators is such that $S(\rho^{(1\,or\,2)})
 \in (0, \log(2s+1)]$. \\

As a consequence of our previous analysis, we can exhibit a unified criterion
 for detecting the entanglement in the boson case.
In order to be unambiguous, such a criterion should make simultaneous
 use of both the Schmidt number and the von Neumann entropy criteria.
In fact, as we have seen before, there exist states with Schmidt number
 equal to $2$, or with von Neumann entropy equal to $1$, which can be
 non-entangled as well as entangled.
Therefore, the only consistent way to overcome this problem derives from
 considering the two criteria together, as clearly stated
 in the next theorem:
\begin{measureboson}
 \label{measureboson}
 A state vector $\vert \psi(1,2) \rangle$ describing two identical
 bosons is {\bf non-entangled} if and only if either its Schmidt number is
 equal to $1$, or the Schmidt number is equal to $2$ {\em and}  the
 von Neumann entropy of
 the one-particle reduced density operator $S(\rho^{(1\: or \:2)})$ is
 equal to $1$.
Alternatively, one might say that the state is {\bf non-entangled} if and
 only if either its von Neumann entropy is equal to $0$, or it is
 equal to $1$ and the Schmidt number is equal to $2$.
\end{measureboson}
%


\section{Conclusions}

The aim of this paper was that of reviewing the delicate problem of
 deciding whether a state describing a system of two identical
 particles is entangled or not.
Following two different, but totally equivalent approaches, we have
 presented two criteria which, in our opinion, should have clarified
 this issue.
The first~\cite{gmw,gm}, in the spirit of the founder fathers of Quantum
 Mechanics, is based on the possibility of attributing a complete set of
 objective properties to both constituents (that is, a definite state vector)
 while the second~\cite{gm2} is based on the consideration of {\em both} the
 Slater-Schmidt number of the fermionic and bosonic analog of the Schmidt
 decompositions of the states describing the system {\em and} of the von
 Neumann entropy of the reduced statistical operators.


\end{document}